\documentstyle[12pt]{article}
\def\ii{\'{\char'20}}

\begin{document}

\begin{titlepage}
\begin{flushright}
{\large UB-ECM-PF-93/10}
\end{flushright}
\title{\large\bf Invariant connections with torsion on group manifolds
and their application in Kaluza-Klein theories}
\author{\large\bf Kubyshin  Yu.A. \\
\thanks{On sabbatical leave from Nuclear Physics Institute, Moscow State
University, 119899 Moscow, Russia}
\thanks{E-mail address: kubyshin@ebubecm1.bitnet} \\
Departament d'Estructura i Constituents de la Mat\'{e}ria \\
Universitat de Barcelona \\
Av. Diagonal 647, 08028 Barcelona, Spain \\  \and
 \large \bf Malyshenko V.O. and Mar\'{\i}n Ricoy D.
\\Nuclear Physics Institute, Moscow State University,\\
Moscow 119899, Russia }
\date{20 April 1993}
\maketitle

\begin{abstract}
Invariant connections with torsion on simple group manifolds $S$ are
studied and an explicit formula describing them is presented. This
result is used for the dimensional reduction in a theory of
multidimensional gravity with curvature squared terms on $M^{4}
\times S$. We calculate the potential of scalar fields, emerging
from extra components of the metric and torsion, and analyze the role
of the torsion for the stability of spontaneous compactification.
\end{abstract}

\end{titlepage}

\section{\bf Introduction}

The ideas of Kaluza and Klein about possible
multidimensional nature  of our space-time were
formulated in the  twenties
\cite{Kaluza} and nowdays
are regarded as an important ingredient of many
supergravity and superstrings theories \cite{sugrav} and
other schemes of unification of interactions.

In the framework of this approach it  is common to assume
that the multidimensional space-time
is of the form $M^{4 } \times I$, where $M^{4}$ is the
macroscopic four-dimensional part of the space-time and $I$
is a compact space of extra dimensions often called
internal space. The compactification of extra dimensions
should occur spontaneously, as a solution of the equations
of motion, which also determine the size $L$ of $I$ after
compactification. In most of the schemes $L\sim L_{Pl}
\sim 10^{-33}$
cm.

In the standard Kaluza-Klein approach the bosonic sector of the
multidimensional theory includes only pure Einstein gravity, and
the electromagnetic and/or non-abelian gauge fields and scalar
fields emerge from the extra components  of the multidimensional
metric after reduction of the theory to the four-dimensional
space-time \cite{kerner}. However, there are some difficulties  in this
approach. One of them is due to the fact that there are no vacuum
solutions with the structure of the space-time $M^{4}\times K/H$,
where $M^{4}$ is the Minkowski space-time and $K/H$ is a homogeneous
space  with non-abelian isometry group $K$, which is necessary for
obtaining non-abelian gauge fields after the dimensional reduction.
Another difficulty is related to impossibility to obtain chiral
fermions in the dimensionally reduced theory. It was found
\cite{Cremmer} that in order to solve these problems one can
generalize the standard Kaluza-Klein approach by adding gauge fields
to the multidimensional Lagrangian (see \cite{salam}, \cite{KMRV} and
\cite{kapetanakis} for reviews). Later other types of generalizations
were proposed. One of them is to remain whithin the pure gravity but to add
terms quadratic in the curvature tensor components.

An interesting possibility in the framework of the latter
generalzation is to consider a model with non-zero torsion.
This option was investigated in a series of papers
\cite {Rudolph} -  \cite{RR}.
There is a big literature on the theory of gravity in the Riemann-Cartan
space-time with torsion (see \cite{HHKN} for a review and an
extensive list of references on the subject).  It is well
known that torsion is a non-dynamical variable in the pure
Einstein - Cartan theory
\cite{Trautmann}, but it becomes dynamical, if one adds quadratic
in curvature terms to the standard Einstein Lagrangian.
Such terms are also motivated by the quantum field theory
limit of strings  \cite{sugrav},
\cite{Schwarz}-\cite{Gross}. Problem of
spontaneous compactification in multidimensional theories
with torsion was investigated in \cite{RR}, \cite{KV} -
\cite{dolan}.

The aim of the present paper is to give a description of
the class of invariant connections with torsion compatible
with the invariant metric on group manifolds $S$, where $S$ is
a simple Lie group, and then apply this result to
investigation of spontaneous compactification in
multidimensional gravity with $R^{2}$-terms on the space-time
$M^{4}\times I$ with $I=S$. We should notice that
description of invariant connections with torsion, in the
case when the internal space $I$ is a homogeneous space
$K/H$ from a certain class, was obtained in \cite{KRR2} and
we will borrow some of the methods from this paper for our
analysis here. But due to some specific features of group
manifolds the case of $I=S$ is not covered by the results
of
\cite{KRR2} and needs a special treatment.

The paper is organized as follows. In section 2 we give a
brief description of $K$-invariant metric compatible linear
connections on homogeneous spaces $K/H$.  In section 3 we
consider the case of group manifolds $S$. We prove a
theorem about decomposition of antisymmetric square of
the adjoint representation of a simple Lie algebra, which
will enable us to construct the invariant connections on $S$
explicitly and to calculate the components of the
multidimensional curvature tensor. In section 4 we
consider multidimensional gravity with $R^{2}$- terms and
derive the potential of scalar fields of the dimensionally
reduced theory. We analyze of this potential
and make conclusions about stability of spontaneous
compactification in the theory.

\section{ General properties of invariant connections on
homogeneous  spaces}

In the present section we present the main results from
the theory of invariant connections on homogeneous spaces
$I=K/H$. They will be used in the next section for explicit construction of
such connections in the case when $I$ has the structure of a simple Lie
group. Our considerations are based on \cite{K-N}.

Let us consider the principle fibre bundle $O(M)$ with the structure
group $SO(d)$, where $d=\dim M$, of
orthonormal frames over $M=K/H$ with reductive
decomposition of the Lie algebra ${\cal K} = Lie (K)$ of the group $K$:
 ${\cal K}={\cal H} \oplus {\cal M}$, $ad H({\cal M}) \subset {\cal M}$,
 where ${\cal H}=
Lie (H)$. The group $K$ acts transitively on the base $M$
and induces a natural automorphism on the bundle $L(M)$. We
will be interested in metrics $g$ on $M$ and metric
compatible connections $\omega$ on the bundle $L(M)$, which
are invariant under the action of the group $K$. The
construction of $K$ - invariant connections on homogeneous
spaces is based on the Wang theorem \cite{K-N}. It states
that there is 1 - 1 correspondence between $K$ - invariant
connections on the bundle $L(M)$ and linear mappings $\Lambda :
{\cal M } \rightarrow {\it so(d)} = Lie (SO(d))$,
which satisfy the following condition
\begin{equation}
\Lambda(Ad h ( X ))  =  Ad  (\lambda ( h ))(\Lambda( X )),\; \; X \in
{\cal M}, h \in H,
\label{eq:equivar}
\end{equation}
where $Ad$ denotes the adjoint representation of $H$ and
$\lambda$ is the homomorphism $\lambda : H \rightarrow SO(d)$ induced
by the action of $H$ on the tangent space $T_{o} (K/H)$
(the corresponding homomorphism of algebras is also denoted
by $\lambda$).  In terms of the mapping $\Lambda$ the
formulas for the invariant torsion and curvature tensors at
the point $o = [H]$ acquire simple form, namely
\begin{eqnarray}
T_{o}( X,Y ) & = & \Lambda ( X ) Y - \Lambda ( Y ) X - [
X,Y ]_{{\cal M}},
\label{eq:torsion} \\
R_{o}( X,Y ) & = & [\Lambda ( X ),\Lambda ( Y )] - \Lambda
( [ X,Y ]_{{\cal M}} ) - \lambda ( [ X,Y ]_{{\cal H}}),\;
\; \; X,Y \in {\cal M}.
\label{eq:curvat}
\end{eqnarray}
Here we identified the tangent space $T_{o}(K/H)$, ${\cal
M}$ and $R^{d}$. Notice also that ${\it so(d)}\cong R^{d} \wedge R^{d}
\cong {\cal M} \wedge {\cal M}$. Let us introduce the mapping
$\beta : {\cal M} \otimes {\cal M} \rightarrow {\cal M}$ by the
formula $\beta ( X,Y ) = \Lambda ( X )Y$. The connection form $\omega$
can be decomposed into the sum of the Levi-Civita connection form
$ \stackrel{0}{\omega}$ with
zero torsion and the contorsion form $\bar{\omega}$.  This
leads to the corresponding decomposition for $\Lambda =
\stackrel{0}{\Lambda} + \bar{\Lambda}$
and $\beta = \stackrel {0}{\beta} + \bar{\beta}$. For reductive
 homogeneous spaces the expression for
$\stackrel{0}{\Lambda}$ was obtained by Nomizu (see \cite{K-N}).
It is given by the formula
\begin{equation}
\stackrel{0}{\beta} ( X,Y ) \equiv \stackrel{0}{\Lambda} ( X )Y = \frac{1}{2}
[ X,Y ]_{{\cal M}} + \stackrel{0}{U} ( X,Y ), \; \; \;  X,Y \in
{\cal M},
\label{eq:levicivita}
\end{equation}
where $\stackrel{0}{U}$ is a symmetric bilinear mapping,
$\stackrel{0}{U} : {\cal M}\otimes{\cal M} \rightarrow
{\cal M}$. We will return to it shortly.

$K$ - invariant metrics $g$ on $K/H$ are known to be in 1 -
1 correspondence with $ad H$ - invariant bilinear forms $B$
on ${\cal M}$, namely $g_{0}( X,Y ) = B ( X,Y ), \; \;  X,Y \in
{\cal M} \cong T_{0}( K/H )$. The invariance of $B$ with
respect to $ad H$ means
\begin{equation}
B ([ A,X ], Y) + B ( X,[ A,Y ]) = 0, \; \; \;  X,Y \in {\cal
M}, A \in {\cal H}.
\nonumber
\end{equation}
It is easy to verify that the condition of compatibility of
the invariant connection form $\omega$ with the invariant
metric $g$ can be written as:
\begin{equation}
B (\beta( X,Y ), Z) + B (Y, \beta( X,Z )) = 0.
\label{eq:compatib}
\end{equation}
We wish to construct the form $\bar{\beta}$, which
describes nonzero torsion.  We can represent it as the sum of
the symmetric and antisymmetric parts: $\bar{\beta} =
\bar{\beta}_{s} +
\bar{\beta}_{as}$  with  $\bar{\beta}_{s}
( X,Y ) = \bar{\beta}_{s}( Y,X )$ and $\bar{\beta}_{as}(
X,Y ) = -
\bar{\beta}_
{as}( Y,X )$. Combining the condition of metric compatibility
(\ref{eq:compatib}) with two other formulas obtained from
(\ref{eq:compatib}) by the cyclic permutation of $X, Y$ and $ Z$ it is easy
to derive the following relation between the
symmetric part ${\beta}_{s}(X,Y) = \stackrel{0} {U}(X,Y) +
\bar{\beta}_{s}(X,Y)$ and
the antisymmetric part ${\beta}_{as}(X,Y) = \frac{1}{2} [ X,Y
]_{\cal M} + \bar{\beta}_
{as}(X,Y)$ of the full mapping $\beta(X,Y)$:
\begin{equation}
B (\beta_{s}( X,Y ), Z) = B ( X,\beta_{as}( Y,Z )) + B( Y,\beta_{as}
(Z,X)), \; \; \; \; X,Y,Z
\in {\cal M}.
\label{eq:equivbeta}
\end{equation}
Thus, if we construct all mappings $\bar{\beta}_{as}$, we will be able to
find all invariant connections on $K/H$ using eq. (\ref{eq:equivbeta}).
Notice that $\bar{\beta}_{as}$ can be
considered as a mapping from ${\cal M} \wedge {\cal M}$
into ${\cal M},\; \bar{\beta}_{as}: {\cal M} \wedge {\cal M}
\rightarrow {\cal M}$.

The condition (\ref{eq:equivar}) can be
rewritten for $\bar{\beta}_{as}$ in the infinitesimal form as
follows:
\begin{equation}
\bar{\beta}_{as} (ad A \wedge {\bf 1} + {\bf 1} \wedge ad A) (\xi ) = ad A
(\bar{\beta}_{as} (\xi)), \; \; \; \xi \in {\cal M} \wedge
{\cal M}, A\in {\cal H}.
\label{eq:condition}
\end{equation}
This enables us to consider $\bar{\beta}_{as}$ as an
intertwining operator, which intertwines equivalent
representations of the algebra ${\cal H}$ in the linear
spaces ${\cal M} \wedge {\cal M}$ and ${\cal M}$. Thus, the
general scheme of construction of the operator $\bar{\beta}_{as}$
is the following \cite{KRR2}.  We decompose linear spaces
${\cal M} \wedge {\cal M}$ and ${\cal M}$ into subspaces
carrying irreducible representations (irreps) of the algebra
${\cal H}$
\[ {\cal M} \wedge {\cal M} = \sum {\cal U}_{k}, \; \; \;
 {\cal M} =
\sum {\cal V}_{n}. \]
According to Schur's lemma, the operator
$\bar{\beta}_{as}$ is equal to $\bar{\beta}_{as} =
\sum f_{kn}{\beta_{kn}}$, where
${\beta_{kn}}$ is the unit operator establishing the isomorphism between the
subspaces  ${\cal U}_{k}$ and
${\cal V}_{n}$ if they carry equivalent irreps and $\beta _{kn} =
0$ otherwise. Similar intertwining operators appear  in the coset
space dimensional reduction of multidimentional Yang-Mills
theories. See \cite{KMRV},  \cite{kapetanakis} for the discussion of
the problem of the construction of such operators in gauge theories.

In order to illustrate the general scheme of calculation of
$\bar{\beta}_{as}$ let us consider two examples.

1. $K/H = G_{2}/ SU(3)$.  From the results in
\cite{37},\cite{Sl} we have after complexification
\begin{eqnarray}
ad {\cal H}({\cal M})& = & \underline{3} \oplus
\underline{3}^{*}, \nonumber  \\
ad {\cal H}({\cal M} \wedge {\cal M})& = & \underline{8}
\oplus
\underline{3}
\oplus \underline{3}^{*} \oplus \underline{1},
\nonumber
\end{eqnarray}
where \underline{8} is the adjoint representation of $ {\cal
H }$. We see that there are only two irreps, $\underline{3}$ and
$\underline{3}^{*}$,
which enter both decompositions.  Therefore, the
intertwining operator is of the form: $\beta_{as} =
f_{33}\beta_{33} + f_{3^{*}3^{*}}\beta_{3^{*}3^{*}}$,
where$f_{33}$ and $f_{3^{*}3^{*}}$ are arbitrary complex
parameters. The reality condition for $\beta_{as}$ implies
$(f_{33})^{*} = f_{3^{*}3^{*}}$.

2. $K/H = ( SU(3) \times SU(3))/ diag ( SU(3) \times SU(3))
\cong
SU(3)$.  In this example
\begin{eqnarray}
ad {\cal H}( {\cal M} \wedge {\cal M} )& = & ad {\cal H}
\oplus
\underline{10}
\oplus \underline{10} ^{*}, \; \;  ad {\cal H} = \underline{8}, \\
ad {\cal H}( {\cal M} ) & = & ad {\cal H }.
\nonumber
\end{eqnarray}
There is only one irrep which enters both decompositions.
Therefore, the intertwining operator has the form
$\bar{\beta}_{as} = f\beta_{88}$, i.e. only one real
contorsion field exists. This example illustrates the case
we are interested in, namely the case
of group manifolds represented as a homogeneous space.

More examples of construction of the contorsion form
as the intertwining operator are given in \cite{KRR2}.

To conclude the section we note that the structure of the
algebra ${\cal K}$ of reductive spaces admits two natural
intertwining operators: $\phi : {\cal M} \wedge {\cal M}
 \rightarrow {\cal M}$ and
$\psi : {\cal M} \wedge {\cal M} \rightarrow {\cal H}$.
They are given by
\begin{eqnarray}
\phi ( X \wedge Y )& = &[ X,Y ]_{{\cal M}}, \label{eq:phi}\\
\psi ( X \wedge Y )& = &[ X,Y ]_{{\cal H}}, \; \; \; X,Y \in {\cal M}.
\label{eq:psi}
\end{eqnarray}
These operators will be used in the next section.

\section { Construction of invariant connections on group  ma\-ni\-folds}

One of the aims of the present paper is to investigate in
detail the case of the group manifolds $S$, represented as
a homogeneous space $S = K/H$ with $K = S \times S$ and $H
= diag(S \times S)$.

There are three natural reductive decompositions for ${\cal K}$
\cite{K-N}, namely
${\cal K} = {\cal H} \oplus {\cal M}$ with ${\cal M} = {\cal M}_{0},
{\cal M}_{+},{\cal M}_{-}$,
\begin{eqnarray}
{\cal M}_{0} &=& \{ ( X/2, -X/2 ), \; \; X \in {\cal S} \},
\nonumber
\\
{\cal M_{+}} &=& \{ ( 0, -X), \; \; X \in {\cal S} \},
\nonumber \\
{\cal M_{-}} &=& \{ ( X, 0), \; \; X \in {\cal S} \},
\label{eq:decomp}
\end{eqnarray}
 with ${\cal H} = Lie (H), {\cal S} = Lie (S)$.
Obviously, $ {\cal M} \cong {\cal H} \cong {\cal S}$.
The $(0)$ - decomposition (the first decomposition in
(\ref{eq:decomp}) ) corresponds to the case when $S$ is represented
as  a symmetric space. The
subspaces ${\cal M}$ and ${\cal H}$ carry the adjoint
representation of $S$ only, therefore, to construct the
mapping $\bar{\beta}_{as}$ in this case we must study the
decomposition of $ad {\cal S} \wedge ad {\cal S}$ into
irreps. The fact, that the $ad {\cal S}$ is contained in $ad
{\cal S} \wedge ad {\cal S}$ at least once is guaranteed by
the existence of the non-trivial intertwining operator
$\phi$ (see (\ref{eq:phi})) in the case of $(\pm)$ -
decomposition and the operator $j \circ
\psi$ (see (\ref{eq:psi})), where $j$ is the isomorphism ${\cal H} \rightarrow
{\cal M}$, for $(0)$ - decomposition.

In fact, we can prove the following

{\bf Theorem.} {\em For simple Lie algebras} ${\cal S}$ {\em
the decomposition of} $ad {\cal S} \wedge ad {\cal S}$ {\em into irreps of}
${\cal S}$ {\em has one of the following forms}
\begin{eqnarray}
ad {\cal S} \wedge ad {\cal S} &=& ad {\cal S} \oplus \gamma
\oplus
\gamma^{*}, \;
\; \; \mbox{for} \; {\cal S} = A_{n}, \nonumber \\
ad {\cal S} \wedge ad {\cal S} &=& ad {\cal S} \oplus
\gamma, \; \; \;  \mbox
{for other simple Lie algebras},
\label{eq:decompS}
\end{eqnarray}
{\em where} $\gamma$ {\em and} $\gamma^{*}$ {\em are irreps
different from} $ad {\cal S}$. {\em Namely}, $\\  \dim \gamma
= n(n-1)(n+2)(n+3)/4$ {\em for} ${\cal S} = A_{n}, \\  \dim \gamma =
n(n-1)(2n-1)(2n+3)/2$ {\em for} ${\cal S} = B_{n} \; \; and \; \;
C_{n}, \\  \dim
\gamma = n(n+1)(2n-1)(2n-3)/2$ {\em for} ${\cal S} = D_{n}, \\
  \dim \gamma = 77$ {\em for}
$G_{2}, 1274$ {\em for} $ F_{4}, 2925$ {\em for} $ E_{6},
8645$ {\em for} $  E_{7}$ {\em and} $30380$ {\em for}
$ E_{8}$.

{\bf Proof.}
For the proof of the theorem it is convenient to complexify ${\cal
M}$ and ${\cal H}$ and apply results from the theory of Lie algebras
\cite{G-G},\cite{Cahn}.  Here we will present the proof for the case
${\cal S} = A_{n}$, other
cases can be proved in a similar way. The main idea is
rather simple. As it has been mentioned above the adjoint
representation $ad {\cal S}$ is contained in $ad {\cal S}
\wedge ad {\cal S}$ at least once. We will find another irrep
$\gamma$ contained in $ad {\cal S} \wedge ad {\cal S}$,
calculate its dimension dim $\gamma$ and show that
\begin{equation}
\dim ( ad A_{n} \wedge ad A_{n} ) = \dim ad A_{n} +
\dim \gamma + \dim \gamma^{*}.
\label{eq:dim}
\end{equation}
We will also see that $ \dim \gamma \neq \dim A_{n}$,
and this will complete the proof of the theorem for ${\cal S} =
A_{n}$. The important tool in the proof is Weil's formula
for the dimension of an irrep
\cite{G-G}. It is known that any irrep $\gamma$ of Lie algebra
${\cal S}$ is characterisized by its highest weight $\Omega = ( \Omega_{1},
\ldots , \Omega_{n})$, where $\Omega_{i}$ are the Dynkin coefficients
of $\Omega$ with respect to a system of the simple roots $\{\alpha_{i}, \;
i = 1, \cdots , n= \mbox{rank} \; S\}$ of ${\cal S}$, i.e. $\Omega =
\sum_{i=1}^{n} \Omega_{i}\alpha_{i}$. Weil's formula states that the
dimension of the irrep $\gamma(\Omega)$ is equal to

\[ \dim \gamma(\Omega) = \sum_{\alpha > 0} \frac{ \sum_{i}
K_{i}(\alpha)  ( 1 + \Omega
_{i} ) \langle \langle \alpha_{i}, \alpha_{i} \rangle \rangle}{ \sum_{i}
K_{i}(\alpha)
\langle \langle \alpha_{i},\alpha_{i} \rangle \rangle}, \]
where the first sum goes over all positive roots
$\alpha_{i}$ of ${\cal S}$ and $K_{i}(\alpha)$ are the coefficients of
the root $\alpha$ with respect to $\{\alpha_{i}\},\;  \alpha = \sum_{i =1}
^{n} K_{i}(\alpha) \alpha_{i}$. Here $\langle \langle \cdot , \cdot
\rangle \rangle$
is the canonical scalar product in the space dual to the Cartan
subalgebra of ${\cal S}$ induced by the non-degenerate  invariant bilinear
form $\langle \cdot , \cdot \rangle $ in ${\cal S}$ ( proportional to the
Killing form ). Nonzero weights of the adjoint representation $ ad {\cal S}$
are roots of the algebra ${\cal S}$.
Our first step is to find any irrep $\gamma$ in the
decomposition of $ad {\cal S} \wedge ad {\cal S}$. There is
a general procedure of finding the so called highest irrep
in the antisymmetric tensor product (see for example ref.\cite{Sl}).
Let us denote by $\Omega_{ad}$ the heighest weight of $ad {\cal S}$ and by
$\tilde{\Omega}_{ad}$ one of the next to the heighest weights, i.e.
the weight which is obtained from $\Omega_{ad}$ by subtraction of one
of the simple roots $\alpha_{i}$ of ${\cal S}$. Then
the highest weight $\Omega$ of the highest irrep in $ad
{\cal S} \wedge ad {\cal S}$ is given by the formula \(
\Omega = \Omega_{ad} + \tilde{\Omega}_{ad}.  \)

For ${\cal S} = A_{n}$ there are two next to the highest weights.
Therefore, two highest irreps $\gamma _{1}$ and $\gamma
_{2}$ in $ad {\cal S} \wedge ad {\cal S}$ exist. Their
highest weights are $\Omega_{1} = ( 0,1,0,\cdots ,0,2 )$
and $\Omega_{2} = ( 2,0, \cdots ,0,1,0 )$. It is known for
${\cal S} = A_{n}$ that if two irreps have the highest weights
$\Omega_{1} = ( a_{1}, \cdots ,a_{n} )$ and $\Omega_{2} = (
a_{n},
\cdots , a_{1} )$, then they are conjugate to each other. Thus, $\gamma
_{2} = \gamma _{1}^{*}$. Weil's formula gives
\[ dim \gamma_{1} = \frac{n(n-1)(n+2)(n+3)}{4} \]
Now we calculate $ \dim {\cal S} = \dim ( ad {\cal S} )$ and $ \dim
( ad {\cal S} \wedge ad {\cal S}):$
\[ \dim (ad
{\cal S} ) = n(n+2), \; \; \;  \dim ( ad {\cal S} \wedge ad {\cal S})
=
\frac{n(n+2)[n(n+2)-1]}{2}. \]
We see immediately that dim $\gamma_{1} \neq \dim (ad {\cal
S})$, therefore $\gamma_{1}$ and $\gamma ^{*}_{1}$ are not
equal to the adjoint representation,
and can check easily that the formula
(\ref{eq:dim}) is true.
This finishes the proof for the case ${\cal S} = A_{n}\; \;  \Box.$

This theorem is the development of the known result stating that
$ad {\cal S}$ is always contained in the tensor product $ad {\cal S}
\otimes  ad {\cal S}$ \cite{Sl}, \cite{Cahn}, \cite{bose}.
Using the theorem proved above we can construct the mapping $\bar{
\beta}_{as}$, satisfying the intertwining condition
(\ref{eq:condition}), explicitly. Indeed, Schur's lemma implies that
$\bar{\beta}_{as}$ must be proportional to the intertwining operator
$\phi$ or $j \circ \psi$ (see (\ref{eq:phi}), (\ref{eq:psi}))
and the result (\ref{eq:decompS}) guarantees that there are no other
intertwining operators mapping from ${\cal M} \wedge {\cal M}$ into
${\cal M}$. Thus, we have
\begin{eqnarray}
\bar{\beta}_{as}( \tilde{X} \wedge \tilde{Y} )  &=& \frac{f}{2} [ \tilde{X},
\tilde{Y} ]_{{\cal M}} \; \; \; \mbox{for $(\pm)$ - decomposition},
\nonumber \\
\bar{\beta}_{as}( \tilde{X} \wedge \tilde{Y} )  &=& 2 f\circ j ([ \tilde{X},
\tilde{Y} ]_{\cal H}) \; \; \; \mbox{for $(0)$ - decomposition},
\label{eq:torsionend}
\end{eqnarray}
where $\tilde{X}, \tilde{Y} \in {\cal M}$ and $f$  is an
arbitrary real parameter.

As for the symmetric part $\beta_{s}$, we can show using
eq.(\ref{eq:equivbeta})  that $\beta_{s}$ is
identically zero. To do this we take an $ad {\cal K}$
- invariant bilinear form $ B (
\tilde{X},\tilde{Y} ) = \langle X_{1},Y_{1}
\rangle + \langle X_{2},Y_{2} \rangle $ on the Lie algebra ${\cal K}$ of $K$.
Here $\tilde{X} = ( X_{1},X_{2} ), \; \; \tilde{Y} = (
Y_{1},Y_{2} ), \; \;
\tilde{X},
\tilde{Y} \in {\cal M}, \; \; X_{i}, Y_{i} \in {\cal S}$, and
as in the proof of the theorem
$\langle \cdot,\cdot
\rangle$ is an $ad {\cal S}$ - invariant bilinear
symmetric form on ${\cal S}
= Lie S$, which in our case is proportional to the bi - invariant
metric $g$ on $S$ and to the Killing form. We see now that
for $\bar{\beta}_{as}$, given by eq. (\ref{eq:torsionend}),  the
r.h.s. of (\ref{eq:equivbeta}) vanishes, thus $\beta_{s} =
0$.

Finally, the mappings $\Lambda$, corresponding to the
invariant connection with torsion on the group manifold
$S$, form a 1 - parameter family given by
\begin{eqnarray}
\Lambda (\tilde{X})\tilde{Y} &=& \frac{1 + f}{2} [ \tilde{X},\tilde{Y} ]_
{{\cal M}}, \; \; \; \mbox{for $( \pm )$ - decomposition},
\nonumber
\\
\Lambda (\tilde{X})\tilde{Y} &=& 2f j([ \tilde{X},\tilde{Y} ])_{{\cal H}},
\; \; \; \mbox{for $( 0 )$ - decomposotion}.
\label{eq:result}
\end{eqnarray}

The mapping $\Lambda$ with $ f=0 $ corresponds to the Levi-Civita
connection with zero torsion on $S$ (see eq.(\ref{eq:result})). When
$f = -1$ for $( \pm )$ - decomposition  and $f = 0$ for $( 0 )$ -
decomposition $\Lambda$ describes the canonical connection. Notice
that for $( 0 )$ - decomposition cononical connection coincides with
the Levi-Civita connection. We would like to underline here that we
constructed non-trivial invariant connection for the case of the $( 0
)$ - decomposition in (\ref{eq:decomp}) when the group manifold $S$
is represented as a symmetric homogeneous space  $K/H$. This differs
from the case of simply connected compact irreducible symmetric
spaces $K/H$ which are not group manifolds. In the latter case, as it
was shown in \cite{KRR2} (~Proposition 3.1), the Levi-Civita
connection is the only $K$ - symmetric metric compatible connection
on $K/H$.

Introducing the isomorphism $i : {\cal S} \rightarrow {\cal
M}$ and using that $\tilde{R} (\tilde{X}_{k},\tilde{X}_{p})
\tilde{X}_{j}$ on
$K/H$ equals $ i(R(X_{k},X_{p}) X_{j}$ on $S$ where $
i(X_{k}) =
\tilde{X}_{k}, \; \; i(X_{p}) = \tilde{X}_{p}$, etc., one
gets from (\ref{eq:curvat})
\begin{equation}
R_{0}(X_{k},X_{p}) X_{j} = F(f) [[X_{k},X_{p}],X_{j}], \;
\; \; F(f)
= \frac{f^{2} - 1}{2}, \; \; \;  X_{k},X_{p},X_{j} \in {\cal S},
\label{eq:fincurv}
\end{equation}
which yields for the curvature tensor components
\[ R_{ijkp} = F(f) C_{kp}^{a} C_{aj}^{b} g(X_{b},X_{i}), \] where
$g( \cdot,\cdot )$ is the bi - invariant metric on $S$ and
$C_{ij}^{k}$ are the structure constants of the algebra
${\cal S}$.

Analogously, eq.(\ref{eq:torsion}) and (\ref{eq:result}) imply that
the torsion tensor on $S$ equals
\begin{equation}
T_{0}( X,Y ) = f[ X,Y ].
\label{eq:fintor}
\end{equation}

\section { Dimensional reduction of multidimensional gravity
with torsion }

We investigate the theory with the action
\begin{equation}
S = \int d\hat{x} \sqrt{ -\hat{g}} \{
\hat{\lambda}_
{0} + \hat{\lambda}_{1}R +
\hat{\lambda}_{2}{\cal
R}^{2} \},
\label{eq:action}
\end{equation}
where ${\cal R}^{2} = \kappa R_{ABCD}R^{CDAB} -
4R_{AB}R^{BA} + R^{2}$, on the space-time $ E = M^{4}\times
S,  \; \kappa = 0,1,2, \; \; \; A,B,C,D = 0,1,2,\ldots ,d+3$. As it
has been pointed out in the introduction such action arises
in the field theory limit of strings with $\kappa = 0$ for
the superstring \cite{Schwarz}, $\kappa = 1$ for the heterotic string
\cite{Gross} and
$\kappa = 2$ for the bosonic strings \cite{Nep}.  We choose the metric
tensor in the block diagonal form
\begin{eqnarray}
\hat{g} = \left( \begin{array}{cc}
g_{\alpha \beta}(x) & 0 \\ 0 &
L^{2}\theta_{a}^{m}(\xi)\theta_{b}^{n}(\xi)\delta_{mn}
\label{eq:metric}
\end{array}  \right),
\end{eqnarray}
where $\alpha,\beta = 0,1,2,3, \; a,b = 4, \ldots, d+3, \;
L$ is a constant of the dimension of length characterizing the size
of the space $S, \; x \in M^{4}, \; \xi \in S, \; \theta_{a}^{m}(\xi)$ are
the vielbeins.  Substituting (\ref{eq:metric}) in
(\ref{eq:action}) and taking the invariance of
the metric and connections into account, we get
\begin{equation}
S = v_{d} \int d^{4}x \; L^{d}\sqrt{- g} \{
\hat{\lambda_{0}} +
\hat{\lambda_{1}}(\bar{R}^{(4)} + R^{(d)}) + \hat{\lambda_{2}}
{\cal R}^{2} \},
\label{eq:action1}
\end{equation}
where $L^{d}v_{d}$ is the volume of the internal space,
$\bar{R}^{(4)}$ and $R^{(d)}$ are the scalar curvatures
of the spaces $M^{4}$ and $S$ respectively, and $g =$ det
$g_{\alpha\beta}$. To separate the term corresponging to the
pure four-dimensional Einstein gravity we introduce the
true physical metric $\eta(x)$ on $M^{4}$ related to $g(x)$
in the following way:
\[ g_{\alpha\beta}(x) = (\frac{L}{L_{0}})^{-d}\eta_{\alpha\beta}(x), \]
where $L_{0}$ is the constant of the dimension of length to be
fixed later on.  Then the action (\ref{eq:action1}) takes the form
\begin{eqnarray}
S = \int d^{4}x \sqrt{-\eta}
\{\bar{\lambda_{1}}R^{(4)} &-&
W(L,f,d,
\kappa,L_{0},\bar{\lambda}_{0},\bar{\lambda}_{2}) \},
\label{eq:ac2}  \\
W(L,f,d,\kappa,L_{0},\bar{\lambda}_{0},\bar{\lambda}_{2}) & = &
- \{
\bar{\lambda}_{2}(\frac{L}{L_{0}})^{d}({\cal R}^{2})^{(4)} + 2\bar{\lambda}
_{2}R^{(4)}R^{(d)}  \nonumber \\
+ \bar{\lambda}_{0}(\frac{L}{L_{0}})^{-d} &+&
 \bar{\lambda}_{1}
(\frac{L}{L_{0}})^{-d}R^{d} +
\bar{\lambda}_{2}(\frac{L}{L_{0}})^{-d}
({\cal R}^{2})^{(d)} \},
\label{eq:pot}
\end{eqnarray}
where $\bar{\lambda}_{i} = \hat{\lambda}_{i}L_{0}^{d}v_{d}$.
If we had considered the contorsion form (\ref{eq:result})
and the metric (\ref{eq:metric}) with the parameter $f(x)$
and the size $L(x)$ depending on the coordinates of
$M^{4}$, then after the dimensional reduction we would have
obtained the Einstein gravity on $M^{4}$ with the metric tensor
$\eta_{\alpha
\beta}(x)$ coupled to the scalar fields $\psi(x) = \ln\{ L(x)/L_{0}\}$
 and $f(x)$ with kinetic
terms, higher derivatives and the potential arising
from (\ref{eq:pot}). If $L$ and $f$ are constant, as in the
case considered in the present paper, we are left with
(\ref{eq:ac2}). Thus, $W$ is the effective
potential of scalar fields $\psi$ and $f$ of the four
dimensional reduced theory. It determines
vacua, i.e. constant with respect to four dimensional coordinates
solutions of the equations of motion. We are
going to analyze the form and properties of the
potential and find its minima.

Assuming that $M^{4}$ is the Minkowski
space-time and $\eta_{\alpha \beta}$ is the Minkowski metric,
the potential $W( L,f,\ldots )$ takes the form
\begin{equation}
W( L,f, \ldots ) = -(\frac{L}{L_{0}})^{d} \{
\bar{\lambda}_{0} +
\bar{\lambda}
_{1}R^{(d)} + \bar{\lambda}_{2}({\cal R}^{2})^{(d)} \} \label{eq:potMink}
\end{equation}
Hereafter we will drop the symbol "$ (d ) $" for the
components corresponding to the space $S$.  The
components of the curvature and Ricci tensors can be expressed in terms
of the eigenvalue of the Casimir operator $C_{2}$.  Using the fact that in
our case $C_{2} = 1$ for the adjoint representation
\cite{kas}
we find for ${\cal R}^{2}$ and $R$
\begin{eqnarray}
{\cal R}^{2} &=& \frac{F^{2}(f)}{L^{4}}d(d+\kappa-4) ,
\label{eq:R2}  \\ R
&=& -\frac{1}{L^{2}}F(f)d.
\label{eq:R}
\end{eqnarray}
Substituting (\ref{eq:R2}), (\ref{eq:R}) into (\ref{eq:potMink}) and
introducing the field $\psi(x) = \ln \{L(x)/L_{0}\}$ we
obtain the following expression for the potential of the
scalar fields $\psi$ and $f$ of the dimensionally reduced
theory
\begin{equation}
W(\psi,f) = e^{-\psi d} \{ \lambda_{0} + \lambda_{1}F
e^{-2\psi }
+
\lambda_{2}F^{2} e^{-4\psi} \},
\label{eq:potMink1}
\end{equation}
where
\[ F(f) = \frac{f^{2}-1}{4}, \; \; \lambda_{0} = -\bar{\lambda}_{0}, \; \;
\lambda_{1} = \bar{\lambda}_{1}\frac{d}{L_{0}^{2}} > 0,\; \; \lambda_{2} =
-\frac{\bar{\lambda}_{2}d(d+\kappa-4)}{L_{0}^{4}}. \]  Our
next step is to investigate possible cases
corresponding to the values of the parameters
$\lambda_{0},\lambda_{1},\lambda_{2}, \kappa$ and $d$ for
which $W(\psi,f)$ has a minimum. We would like to underline
here that the point $(\psi_{min},f_{min})$, where the
potential has its minimum, is a constant solution of the
equations of motion of the theory corresponding to
spontaneous compactification of the extra dimentions to the
compact space $S$, so that the space-time has the form
$M^{4}\times S$ (see \cite{VK}).
We will use the
notation $\Delta \equiv
\lambda_{1}^{2}/4\lambda
_{0}\lambda_{2}$.

\subsection{ Case {\rm 1}: $\; \lambda_{0},\lambda_{2} >0, \; \Delta = 1$ }

It can be checked that in this case the potential has
degenerate minima at the points $(\psi_{min},f_{min})$
located on the curve \( \psi_{min}(f) = \frac{1}{2} \ln\{
\lambda_{2}(1-f^{2})/2\lambda_{1}\} \), $\mid f \mid
< 1$, and its values at these points are equal to zero, i.e.
$W(\psi_{min},f_{min}) = 0$ (see Fig. 1). The four-dimensional
cosmological constant $\Lambda^{4}$, which is determined by
the value of $W$ at the point corresponding to the vacuum
solution, vanishes.  We may fix the parameter $L_{0}$ by the
requirement $\psi_{min}( 0 ) = 0$. This gives
\[ L_{0} = \sqrt{\frac{-\hat{\lambda}_{2}(d+\kappa-4)}{2\hat
{\lambda}_{1}}}, \]
which is thus the size of the internal
space in the vacuum corresponding to spontaneous
compactification with zero torsion.

\subsection{ Case {\rm 2}:  $ \; \lambda_{0},\lambda_{2} >0, \; \frac{d(d+4)}
{(d+2)^{2}} < \Delta <1$ }

In this case the potential is
positive for all $(\psi,f)$ and vanishes when $\psi
\rightarrow +\infty$. It can
be verified that the potential has the minimum at the point
\[ (\psi_{min},f_{min}) = (\frac{1}{2} \ln\frac{\lambda_{2}(d+4)}
{2\lambda_{1}(d+2)[1+\sqrt{1+\frac{d(d+4)}{(d+2)^{2}}\Delta^{-1}}]},0 ). \]
Since $W(\psi_{min},0) >0$ the mi\-ni\-mum is local and the
cor\-res\-pond\-ing vac\-uum state is me\-ta\-sta\-ble. We should
note that the potential $W(\psi,f)$ has two gutters in the
region $\mid f \mid <1$ and $\psi < \frac{1}{2}
\ln{ 2\lambda_{1}/ \lambda_{2}}$,
which join each other at the point $(\frac{1}{2} \ln{
2\lambda_{1}/\lambda_{2}} ,0)$ and ascend when $\psi \rightarrow
-\infty$. We fix the parameter $L_{0}$ by the requirement
that $\psi _{min} = 0$.
This gives two values for $L_{0}^{2}$ and we choose the one
for which $L_{0}^{2} = 0$ when $\lambda_{2} = 0$ ( this
corresponds to the collapse of the internal space and
absence of stable spontaneous compactification solution
for the pure Einstein gravity, as it has been discussed in the
Introduction):
\begin{equation}
L_{0}^{2} =
\frac{(d+2)\hat{\lambda_{1}}}{2\hat
{\lambda}_{0}}[-1
+\sqrt{1-\frac{\hat{\lambda}_{0}\hat{\lambda
_{2}}}{\hat{\lambda}_{1}^{2}}(d+\kappa-4)(d-4)} \; ].
\label{eq:size}
\end{equation}
The experimental bound for the four-dimensional cosmological
constant $\Lambda^{4}$ gives $\mid 16\pi G\Lambda^{4} \mid
< 10^{-120}$. By making the parameter $\lambda_{0}$
approaching $\lambda_{1}^{2}/4\lambda_{2}$ from
above we can obtain arbitrary small values for $\Lambda^{4}
= W(0,0)$.  Trying to make $\Lambda^{4} = 0$ it is easy to see
 that this is possible if and only if $\Delta = 1$,
i.e. in case {\rm 1}.

\subsection{ Case {\rm 3}: $ \; \lambda_{0},\lambda_{2} >0, \; \Delta >1$ }
Now the potential has no minimum. As in the previous case for
$\psi < \frac{1}{2} \ln {2\lambda_{1}/\lambda_{2}}$ and
$\mid f \mid <1$, the potential has two gutters which join
each other at $( \frac{1}{2}\ln{2\lambda_{1}/\lambda_{2}},
0)$, but contrary to the case {\rm 2}, lower to $(-\infty)$ when
$\psi \rightarrow -\infty $. These features of the
potential are of some importance for understanding of the
spontaneous compactification issue as will be discussed in the
next section.

\section {\bf Discussion of the results }

Let us start our discussions with case {\rm 2}. We have found that the
minimum of the scalar potential
$W(\psi,f)$ of the reduced theory is at $(\psi_{min},0)$ (for our choice of
$L_{0} \; \;  \psi_{min} = 0)$ that corresponds to spontaneous
compactification of extra dimensions to the group
manifold $S$ with characteristic size $L_{0}$ given by
(\ref{eq:size}) and zero torsion. This minimum is stable
with respect to fluctuation with non zero torsion and
classicaly stable with respect to fluctuation in $\psi$ -
direction. Since $W(0,0) >0$ and $W(\infty,0) = 0$ the corresponding
vacuum state is metastable, and the system can pass
to the region of large $\psi$ via quantum tunnelling.  This
corresponds to decompactification of the space of extra
dimensions. Analogous phenomena for the multidimensional
Einstein - Yang - Mills system without torsion were
considered in \cite{KRT},\cite{BKM}. But if the parameters
of the lagrangian are tuned in such
way that the four-dimensional cosmological constant
$\Lambda^{4} = W(0,0)$ is small enough to satisfy the
experimental bound $\mid 16 \pi G\Lambda^{4} \mid <
10^{-120}$, the lifetime of the metastable vacuum exceeds
the lifetime of the Universe (see estimations in
\cite{BKM}).

Another interesting problem which can be addressed here is
the dynamics of compactification of extra dimensions in
the framework of the Kaluza - Klein cosmology (the
analogous issue for the Einstein - Yang - Mills system
without torsion was considered in \cite{KRT}).  The main
question is the following: if at the early stage the
multidimensional Universe had started its classical
evolution with large negative $\psi$ (small size $L$) and
large $\mid \dot{\psi} \mid$, would it have found
itself in the minimum $( \psi = 0,f= 0)$ corresponding to
spontaneous compactification of extra dimensions? Since the
height of the barrier separating the minimum from the
region where $\psi \rightarrow +\infty$ (decompactification
of extra dimensions ) is finite, the system can have
enough energy to overcome the barrier in spite of the
loss of energy due to the friction terms which are present in
such sort of theory. In any case this question needs more
detailed investigation for which the explicit form of
non-static terms in the reduced action must be known. This
is beyond the scope of the present paper.

In case {\rm1}, the minimum of the potential is degenerate:
$W(\psi_{min}(f),f) = 0$ for $0\leq \mid f \mid \leq 1$.
The vacuum with $(\psi_{min}(0),0)$, corresponding to
compactification with zero torsion, is not separated by any
barrier from another vacua with the same energy but non-zero torsion.

The situation changes even more drastically in case {\rm3}. The
potential $W(\psi,f)$ does not have any minimum at all. But
if we analyzed the same theory without torsion, we would
see (as in case \rm{1} also) that the potential $W(\psi ,0)$ has
minimum and expect to have spontaneous compactification
solution. However, taking torsion as additional degree of
freedom in the theory into account changes the situation.
The vacuum $(\psi_{min}(0),0)$ is not stable and
small fluctuations of the fields and their time derivatives may
initiate transitions of the system to another state with the same
(case {\rm1}) or less (case {\rm3}) energy. Non-zero torsion is developed in
such transitions.

We think that this example is rather instructive for deeper
understanding of the spontaneous compactification problem. It
illustrates some of the hidden difficulties that the Kaluza
- Klein approach may face.

In conclusion we would like to underline that the
analysis of the Kaluza - Klein ${\cal R}^{2}$ - gravity
with torsion was carried out for arbitrary $S \times S$ - invariant
configurations of the metric and connection form. The mathematical
results, obtained in Section 3, enabled
us to describe all metric compatible invariant connections
with non-zero torsion on group manifolds $S$, and thus the
solution of the spontaneous compactification problem for
this class of metrics and connections is complete.

\section {\bf Acknowledgements}

It is a pleasure for us to thank A.P. Demichev A.P., J.I. P\'{e}rez Cadenas,
G. Rudolph G. and I.P. Volobujev I.P. for their
useful comments and critical remarks and J. Mour\~{a}o for illuminating
discussions and critical reading of the manuscript. Yu. K.
acknowledges support from Direcci\'{o}n General De Investigaci\'{o}n
Cient{\ii}fica y T\'{e}cnica (sabbatical grant SAB 92 0267) during part of
this work and thanks the Department d'ECM de la Universidad de Barcelona
for its warm hospitality.

\newpage
\begin{description}
\item[{\bf Figure 1}] Shape of the potential $W(\psi, f)$ when $\lambda_{0},
           \lambda_{0} > 0$ and $\Delta = 1$ (see Sect. 4.1). The minimum
    of the potential is degenerate and is located on the curve
    $\psi - \frac{1}{2} \ln \frac{(1-f^{2})}{2 \lambda_{1}} = 0$
    in the $(\psi, f)$-plane.
\end{description}

\end{document}